\documentstyle[aps,prl,epsf]{revtex}
\begin{document}
 
\twocolumn[\hsize\textwidth\columnwidth\hsize\csname
@twocolumnfalse\endcsname

\title{A model for the fragile-to-strong transition in water}
\author{E. A. Jagla}
\address{The Abdus Salam International Centre for Theoretical
Physics (ICTP), I-34014 Trieste, Italy}
\maketitle

\begin{abstract}
A model based on the existence of two different competing local
structures in water is described. It is shown that it can explain the
transition between fragile and strong behavior that supercooled water has around
220 K. The high temperature behavior is similar to that 
observed in standard fragile glass-formers. The strong
behavior at low temperatures is associated with the existence of a remanent
configurational
entropy coming from the possibility of locally  choosing  between the two
possible environments.

\end{abstract}
 
\vskip2pc] \narrowtext

\section{Introduction}

The form of the dependence of viscosity on temperature is 
among the many properties that make water an anomalous fluid.
Water is a fragile fluid when viewed at
temperatures close to
the melting temperature, indicating that there is an arrest of its
degrees of freedom on cooling. This behavior is typical
of many substance known as fragile glass-formers
\cite{angell91,angell95}. However, close to the
glass temperature $T_G$ ($\sim 136$ K)supercooled water shows characteristics of a
strong
liquid \cite{ito99,sastry99}, in which there is an almost temperature
independent configurational entropy,
that manifests in an Arrhenius dependence of the viscosity $\eta$ as a
function of $T$.
Thermodynamic constraints limit the transition between these  
two regimes to occur rather sharply in a temperature range around
$\sim 220$ K\cite{starr99}.

There is by now a good piece of evidence that many of the anomalous properties
of water can be rationalized by the use of an effective, two particle,
spherical interaction potential, of the core-softened
type\cite{eajpre,eajjcp2,sadr,speedy97}. This
interaction can be viewed as appearing between clusters of water molecules,
rather than between single molecules\cite{stanleyrev,campolat}. The main
characteristic of this
interaction is that it allows for two different equilibrium distances
between clusters, depending on pressure. An appropriate, simplified model
that capture many of the anomalies of water is
provided by
spherical particles interacting through a potential consisting of a
hard core plus a soft repulsive shoulder \cite{eajpre,eajjcp2}. 
Here we show--using an analytically solvable version of it--that this kind of
interaction can also explain the non-standard
behavior of $\eta(T)$.

\section{Hard spheres model}

We will use a model of hard spheres as a starting point (in the next section
it will be generalized to describe the properties of water).
We will suppose that the pure hard sphere system has an
ideal
thermodynamical glass transition at some temperature $T_0$ when the fluid
phase 
is supercooled, preventing crystallization. A possible scenario for this
glass transition is the following\cite{speedyhs0,speedyhs1,speedyhs2}.
For glassy systems there is a contribution $s_c$ to the entropy--referred
to as
configurational entropy--that comes from the many different configurations in
which the glass can exist. For the case of spheres it comes
from the many ways
in which the spheres can be accommodated in stable, non-crystalline arrangements.
These configurations differ in the value of the specific volume $v$. 
We will suppose that hard spheres have a configurational entropy per particle
$s_c^{HS}(v)$ of the form\footnote{In Ref. \cite{speedyhs2} a
parabolic form for
$s_c$ as a function of {\em density} (instead of $v$) is used. The
difference between
both choices is tiny, and both give rise to an ideal glass transition.}
\begin{equation}
s_c^{HS}(v)=\alpha(v-v_0)-\beta(v-v_0)^2.
\end{equation}
According to this formula, $s_c^{HS}$ becomes lower than zero for $v<v_0$ (and
$v>v_0+\alpha/\beta$), indicating
that there are no accessible states in this range, i.e., $v_0$ is the minimum
value that $v$ can take. To get the total entropy $s_{\rm tot}$ of
the system we should still include the contribution coming from small 
vibrations around each configuration \cite{speedyhs2}. However, for our
purpose this will not be necessary, since (as we will see below) $\eta$ is
given in terms of $s_c$, rather than $s_{\rm tot}$, and $s_c(T)$ is
independent of the vibrational contribution to the entropy.
The previous form of the configurational entropy
implies the existence of an ideal thermodynamical glass transition occurring
at $T_0$, where $T_0$ is obtained from
$P/T_0=\left .\partial
s_c^{HS}/\partial v\right |_{v=v_0}=\alpha$. Whereas from microscopic grounds
there
is no rigorous prove that this transition should occur, the consequences on
observable magnitudes that can be predicted from it are consistent with the
known phenomenology of glassy systems\cite{young} and with results of numerical
simulations\cite{speedyhs2,speedy-deb}.
For $T<T_0$ the system is in the
fundamental configurational state, and then $s_c^{HS}(T<T_0)=0$. For
$T>T_0$,
$s_c^{HS}$
is given by
\begin{equation}
s_c^{HS}(T>T_0)=\frac{\alpha^2}{4\beta T^2}(T^2-P^2/\alpha^2)=
\label{schs}
\frac{\alpha^2}{4\beta T^2}(T^2-T_0^2)
\end{equation}
This expression for $s_c^{HS}$ can be used to calculate transport
properties such
as the viscosity $\eta$ through the use of the Adam-Gibbs formula
\cite{adam65}.
It states that the value of $\eta$ is given by
\begin{equation}
\eta(T)=\eta_0\exp[A/(Ts_c)],
\label{ag}
\end{equation}
which for hard spheres becomes
\begin{equation}
\eta^{HS}(T)
=\eta_0\exp\left [4A\alpha ^{-2}
\beta T/(T^2-T_0^2) \right]
\label{ag2}
\end{equation}
where $\eta_0$ and $A$ are constants.
This is a behavior typical of a fragile glass-former, in which $\eta$
increases
more rapidly than in a simple thermally activated process, and it diverges
when $T\rightarrow T_0$. The presence of $s_c$ in (\ref{ag}) reflects the fact that
jumps between different basins of the energy landscape become less probable as the
number of these basins diminishes.

\section{Core-softened models for water}

Properties of water have been studied recently by using models in which
particles interact through potentials that allow for two different equilibrium
distances between particles, namely $d_0$ and
$d_1>d_0$\cite{eajpre,eajjcp2,sadr,cho96}. One possibility is, for
instance, to take a strict hard core at $d_0$ and a shoulder that vanishes at
$d_1$\cite{eajpre,eajjcp2}. Here we will use a further
simplification of this kind of models in
order to be able to extract analytical results\cite{eajjcp2}.
We consider spheres of radius $r_1$ ($=d_1/2$). Pairs of spheres will be
allowed to overlap (more than two overlapping spheres will not be
allowed), and each time this happens the system will be charged an energy
$\varepsilon_0$. This may be considered as a limiting case of particles with
a core-softened potential, in which there is a low energy hard core at a
distance $2r_1$, and a strict hard core at $2r_0$, and we are 
taking $r_0=0$.
To make the problem analytically tractable, we
will also suppose that each time two spheres overlap, they are constrained  
to have their centers in exactly the same position. This approximation
neglects the entropy associated with small vibrations of the particles in each
pair.

We are interested in the configurational entropy $s_c$ of the system, now as a
function of the specific enthalpy  $h=Pv+e$, that includes the
internal energy $e$ coming from the existence of overlapping particles. To
calculate $s_c(h)$ we proceed in
the following way. Suppose we have a system of $N$ particles, $n$ of them in
non overlapped positions  and $n'$ pairs of overlapped particles
($N=n+2n'$). The configurational entropy will be that of $n+n'$ hard spheres
plus the combinatorial entropy for locating the $n'$ pairs in the $n+n'$
possible positions, i.e.,
\begin{equation}
\tilde s_c=\frac{n+n'}{N}s_c^{HS}+ k_B\ln \left (\begin{array}{c}n+n' \\
n'
\end{array}
\right )
\end{equation}
(here we use $\tilde s_c$ to indicate an entropy functional).
Using $x\equiv n'/N$ and  $\tilde v \equiv V/(n+n')=v/(1-x)$ as
independent variables we can
write $\tilde s_c$ as a

\begin{eqnarray}
\tilde s_c(x,\tilde v)&=&(1-x)s_c^{HS}(\tilde v)+ \nonumber \\
&+&k_B \left [ (1-2x)\ln \left ( \frac{1-x}{1-2x}\right )
+x \ln\left( \frac{1-x}{x}\right )\right ].
\end{eqnarray}
In order to get the thermodynamic value of $s_c(h)$ at each value of the
external pressure $P$, we have to maximize $\tilde s_c$ for each fixed
value of the enthalpy, i.e., 
\begin{eqnarray}
s_c(h)&=&\left .\min_{x,\tilde v}\right |_{h} \tilde s_c(x,\tilde v)\\
h=Pv+e&=&(1-x)P\tilde v+x\varepsilon _0.\label{hache}
\end{eqnarray}

\section{Results}

In Fig. \ref{f1} we see $s_c$, $x$, and $v$ as functions of $h$ for three
different values of
$P$, using expression (\ref{schs}) for $s^{HS}(v)$  with $\alpha =2.79~ k_Br_1^{-3}$,
$\beta =0.97~ k_Br_1^{-6}$, and $v_0=6.37~r_1^3$ which are values extracted from
numerical 
simulations of hard spheres systems \cite{speedyhs2}. We also plot in Fig.
\ref{f1} the
limiting cases $\tilde s_c(x=0)$ and $\tilde s_c(x=0.5)$, corresponding to all
particles in singled or overlapped positions, respectively. The thermodynamic
value $s_c(h)$ can never be lower than $\tilde s_c(x=0)$ or
$\tilde s_c(x=0.5)$. The states with the lowest enthalpy for $x=0$ and
$x=0.5$ have (from (\ref{hache})) $h=Pv_0$ and $h=Pv_0/2+\varepsilon_0/2$,
respectively. These values coincide at $P_{\rm cr}\equiv\varepsilon_0/v_0$.
In Fig. \ref{f1}(a), $P=0.9~P_{\rm cr}$, $\tilde s_c(x=0)$ is always
greater than $\tilde s_c(x=0.5)$
and for this reason $x$ takes 
values close to 0, indicating that most particles are in singled
positions. The obtained $s_c(h)$ function departs from zero with
infinite derivative at $h=0.9~\varepsilon_0$ (because of the combinatorial
contribution
to the entropy) but it still has a singularity (namely a jump in its second
derivative) when $\partial s_c/\partial h=\alpha/P$.
In (b) the value of $P=1.1~P_{\rm cr}$ is larger, and at low $h$ the
contributions
with
$x=0.5$ dominate, indicating that the system has almost 
all particles coupled in pairs. For higher $h$, $x$ goes down to zero, namely
paired particles become rare. In (c), the value of $P=P_{\rm cr}$ is
exactly that
at which the ground state of the system  with $x=0$ and the one with
$x=0.5$ are degenerated. In this case, entropy starts from a finite
value $\simeq 0.48~k_B$ at $h_{\rm min}=\varepsilon_0$, corresponding to
the maximum
combinatorial entropy of choosing which particles are singled, an which are
paired.

To get the values of the thermodynamic variables as a function of $T$,
instead of $h$, we have only to make use of the relation $T^{-1}=\partial
s/\partial h$. The results for the configurational entropy $s_c(T)$ and
the viscosity $\eta(T)$ (calculated using the Adam-Gibbs formula (\ref{ag})) are
shown in Fig. \ref{f2}.

\section{Discussion and comparison with water}

\begin{figure}
\narrowtext
\epsfxsize=3.3truein
\vbox{\hskip 0.05truein
\epsffile{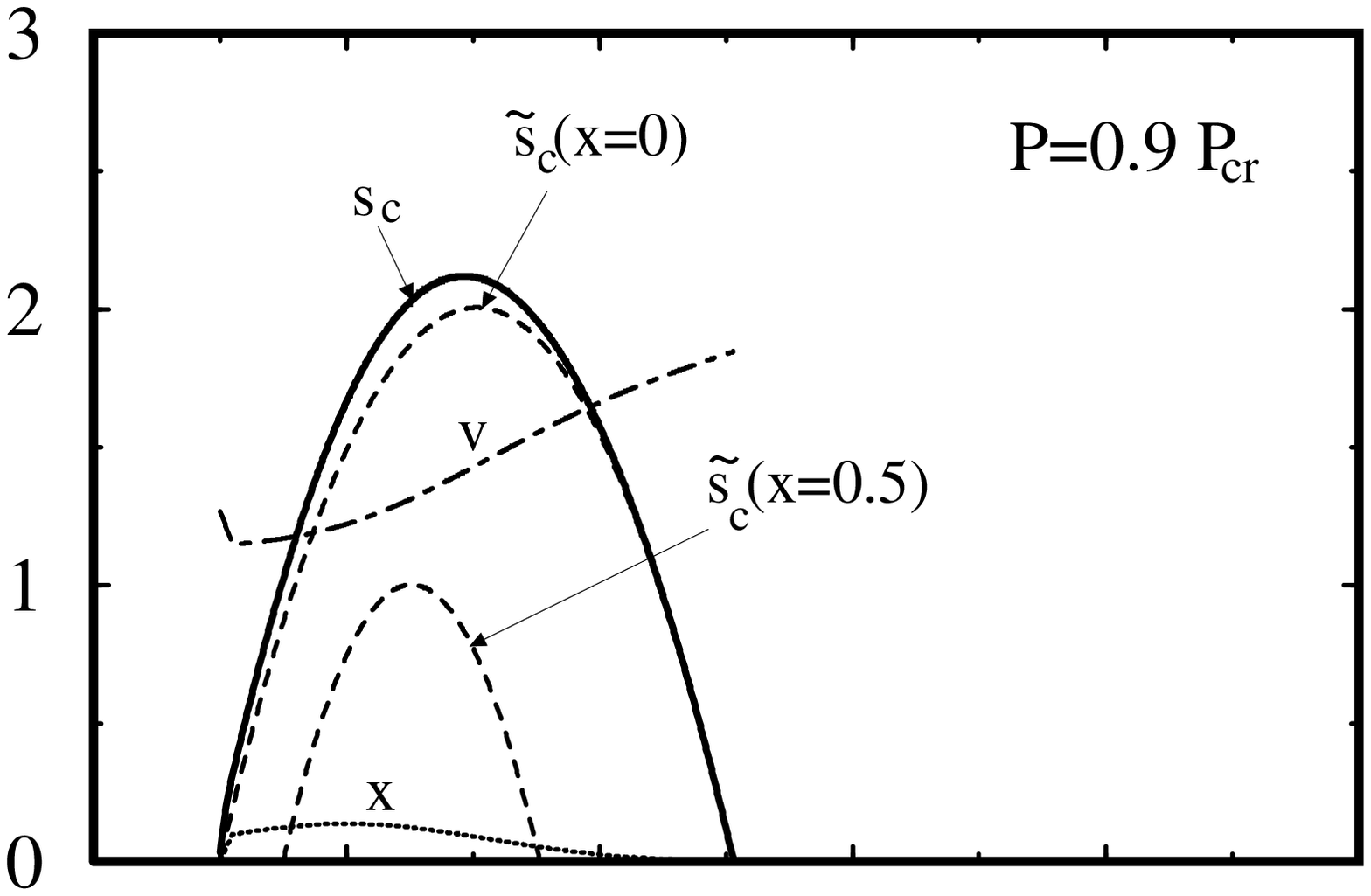}}
\epsfxsize=3.3truein
\vspace{-2.6cm}
\vbox{\hskip 0.05truein
\epsffile{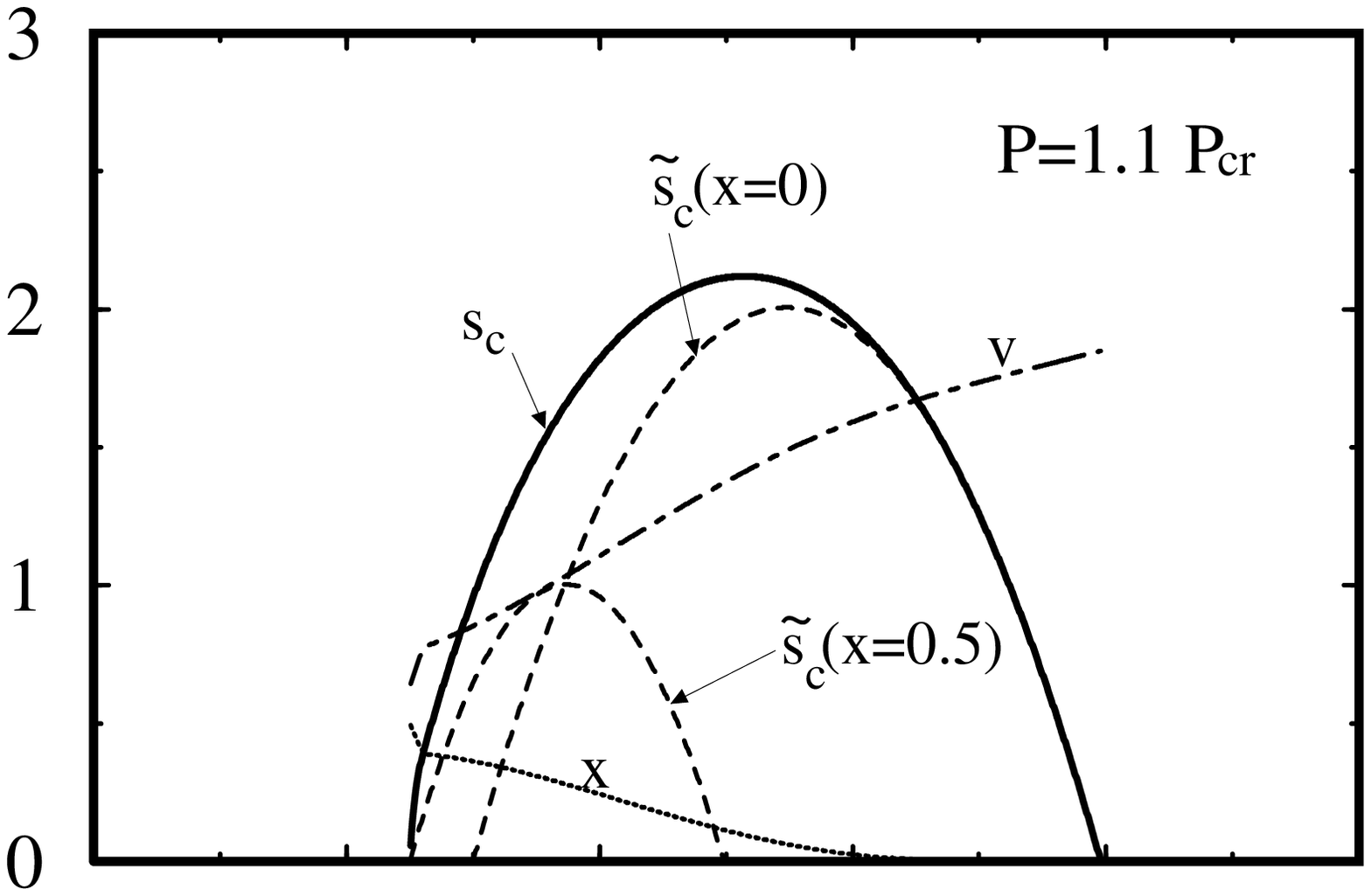}}
\epsfxsize=3.3truein
\vspace{-2.6cm}
\vbox{\hskip 0.05truein
\epsffile{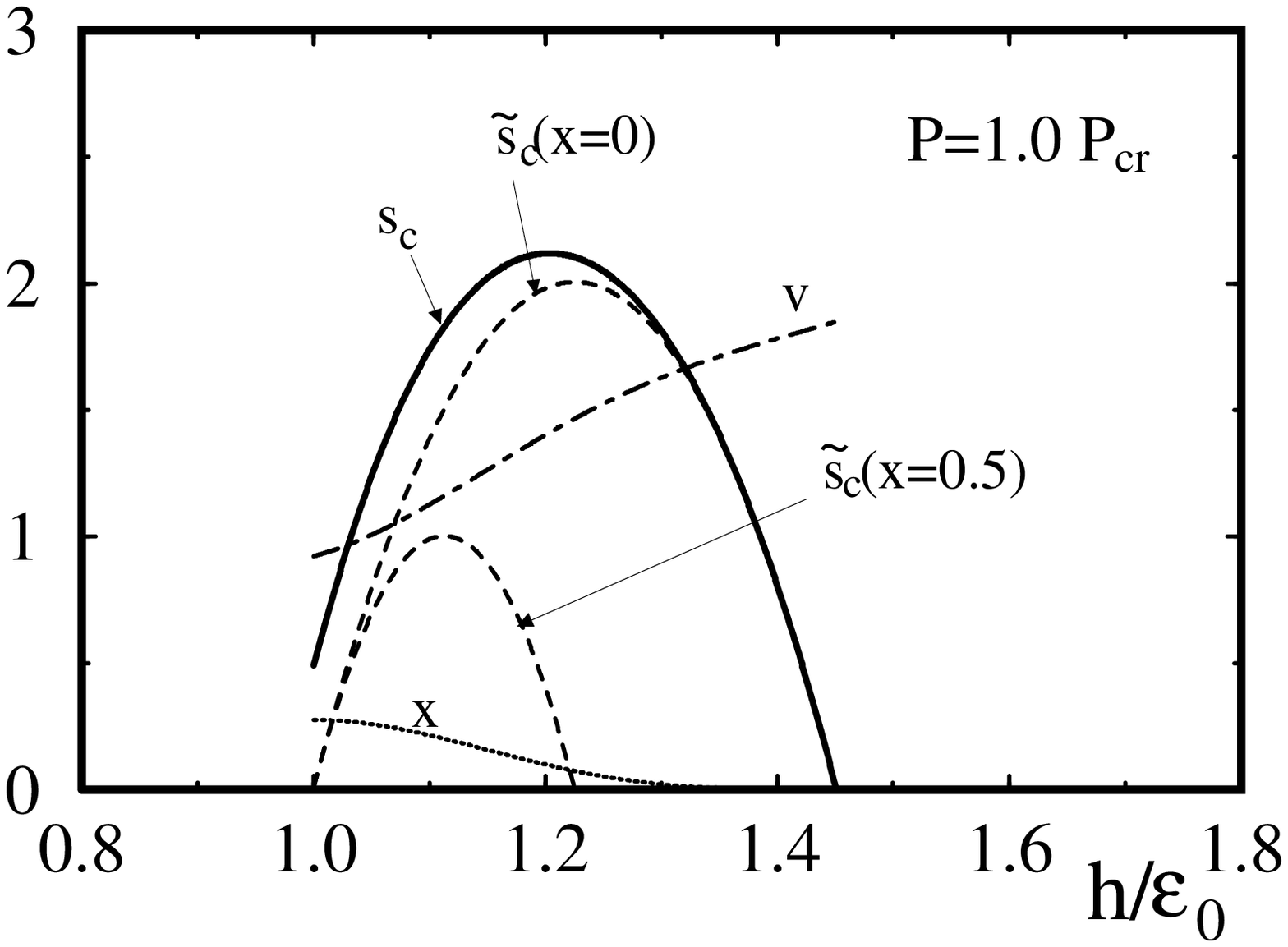}}
\medskip
\caption{Configurational entropy $s_c$, specific volume $v$, and fraction of pairs of
particles relative to the total number of particles $x$ as a function of the enthalpy
for three different values of pressure.
In order to compare, the limiting
cases $\tilde s_c(x=0)$ and $\tilde s_c(x=0.5)$ are also shown (entropies are given in 
units of $k_B$, $v$ is in  
units of $5r_1^3$, where $r_1$ is the radius of the particles, see text for details).}
\label{f1}
\end{figure}

\begin{figure}
\narrowtext
\epsfxsize=3.3truein
\vbox{\hskip 0.05truein
\epsffile{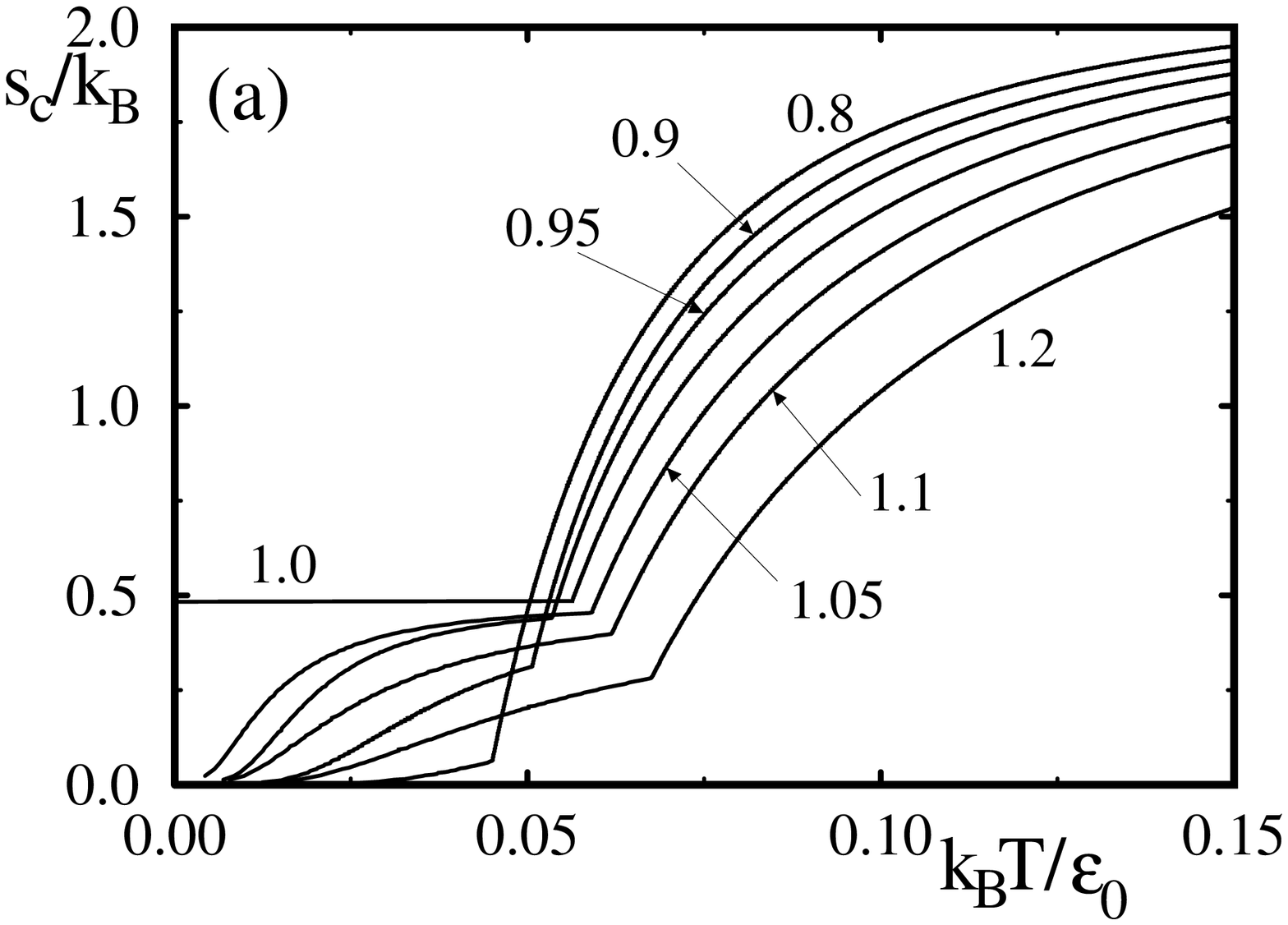}}
\epsfxsize=3.3truein
\vspace{-1.5cm}
\vbox{\hskip 0.05truein
\epsffile{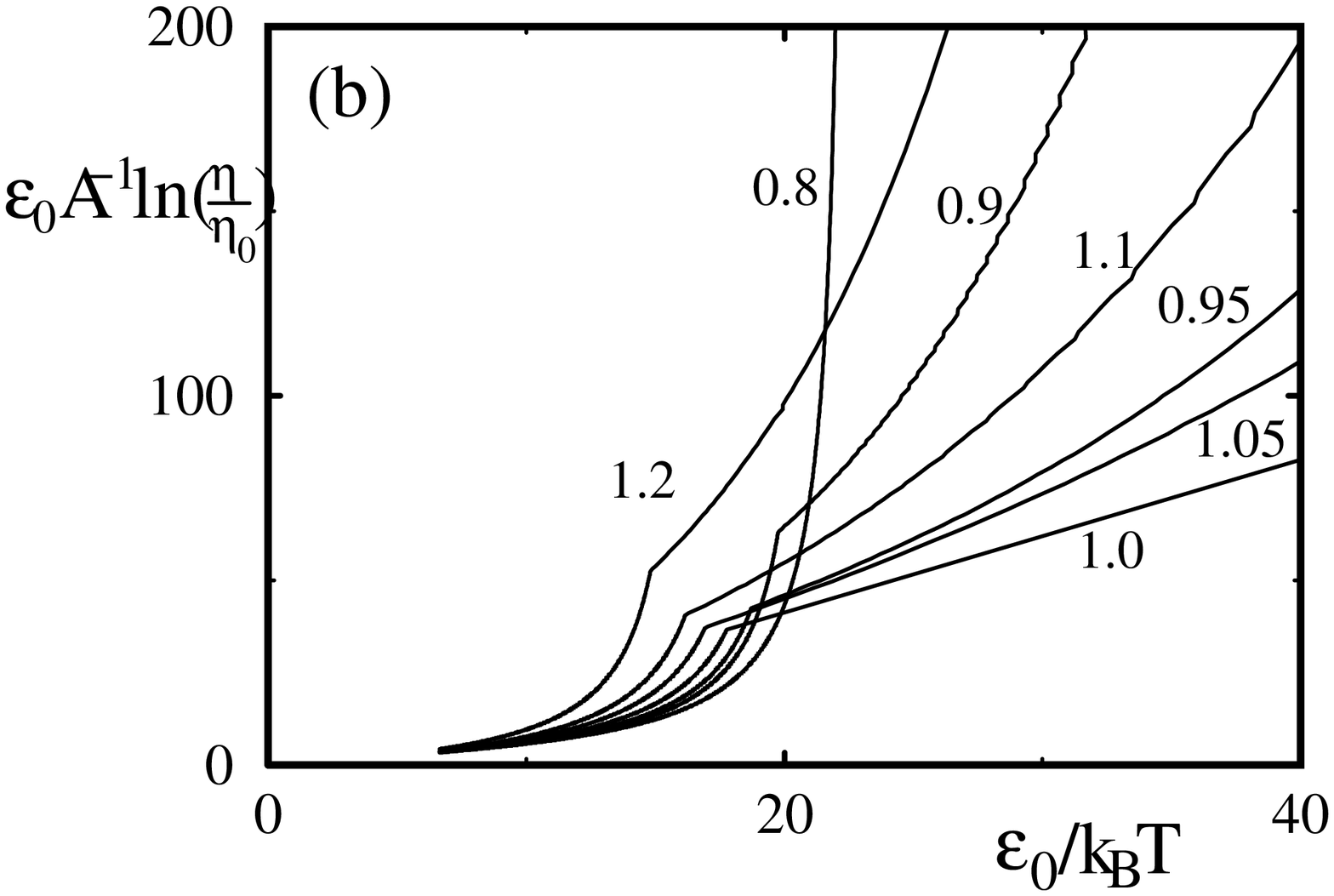}}
\medskip
\caption{(a) Configurational entropy $s_c$ as a function of $T$, for
different values of
$P/P_{\rm cr}$, as indicated. (b) Viscosity $\eta$ vs. $1/T$, calculated according to
the Adam-Gibbs formula (expression (\ref{ag}), $\eta_0$ and $A$ are the constants in
that expression).}  
\label{f2}
\end{figure}

From Fig. \ref{f2}(a) we see that for any $P$, $s_c$
is finite for all $T\neq 0$, i.e., there is no vanishing of $s_c$ at any
finite 
temperature, contrary to what happened in the case of simple hard spheres 
(see Eq. (\ref{schs})).
This is due to the possibility for the system of having
particles singled or paired, which always accounts for the existence 
of a non-zero combinatorial entropy. For $P=P_{\rm cr}$ the bottoms of the
$\tilde s_c$ functions
corresponding to $x=0$ and $x=0.5$ coincide (Fig. \ref{f1}(c)),
and this combinatorial entropy can be
used up to $T=0$, in such a way that  $s_c$ remains finite, even  when
$T\rightarrow
0$. For $P\neq P_{\rm cr}$, $s_c$ goes to zero as $T\rightarrow 0$, as the
ground state
is unique. There is still a phase transition at finite temperatures,
signaled by the kink in the $s_c(T)$ curves in Fig \ref{f2}(a). The position of
this kink in the $P$-$T$ plane is given by $P/T=\alpha$.
In the $\log(\eta)$-$1/T$ plot (Fig. \ref{f2}(b)), the kinks mark the
transition
between a fragile and a strong behavior. At high $T$, $s_c$
diminishes rapidly with temperature and the system is fragile. This
behavior is equivalent to that
of the simple hard sphere system (see Eq. (\ref{ag2})). At low $T$, the
dependence of $s_c$ on $T$ is much weaker, indicating a stronger behavior. 
In this regime, the available configurational entropy is mainly of
combinatorial nature.

\begin{figure}
\narrowtext
\epsfxsize=3.3truein
\vspace{-1.5cm}
\vbox{\hskip 0.05truein
\epsffile{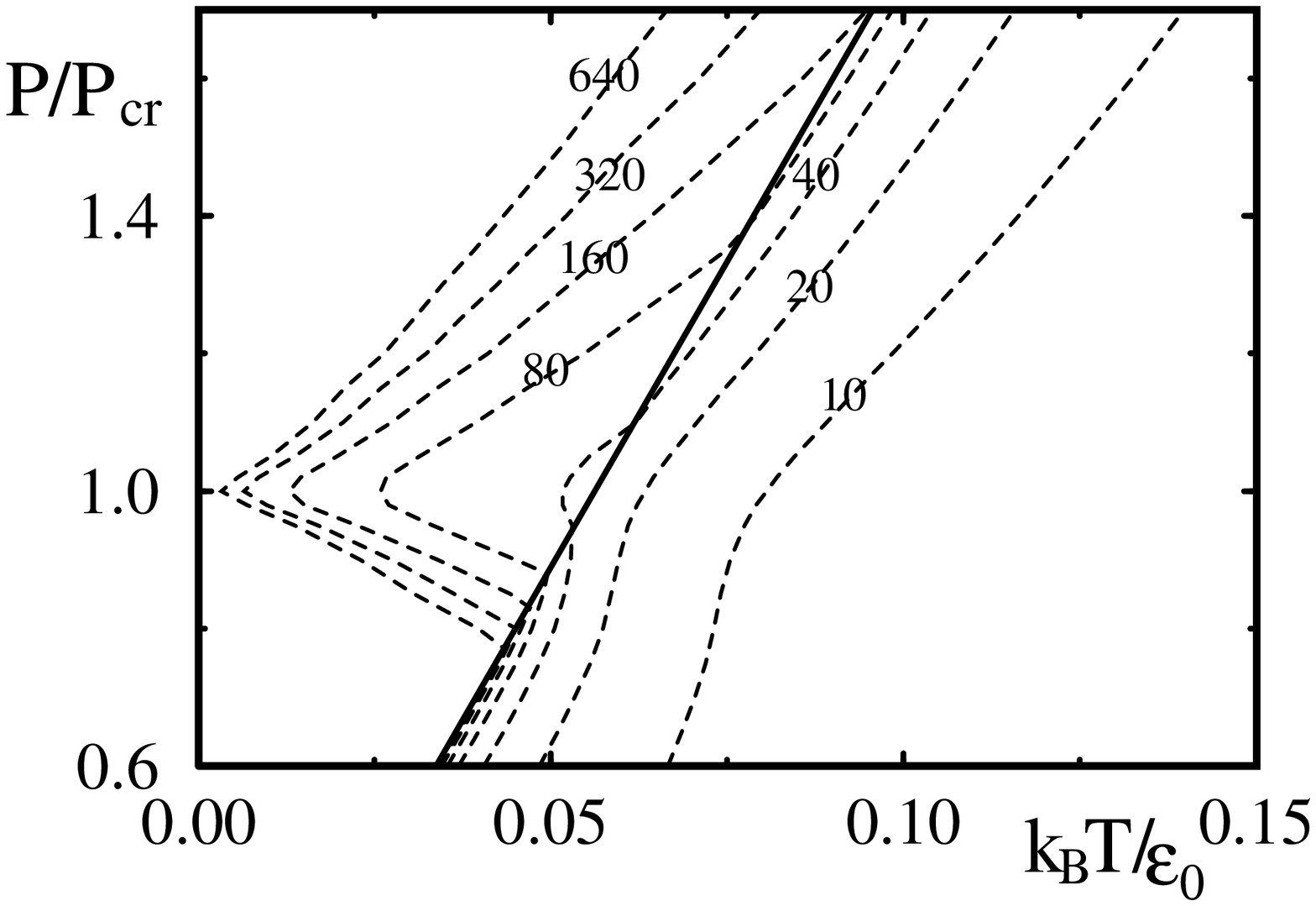}}
\medskip
\caption{Pressure-temperature plot, indicating the position of the fragile-to-strong
transition (solid line) and contour lines of constant viscosity (dashed lines, the
corresponding values of $\varepsilon _0A^{-1}\ln (\eta/\eta_0)$ are
indicated).} 
\label{f3}
\end{figure}

For this model there is no ideal glass transition, i.e., $\eta$ is finite at
any finite temperature, and $T_0=0$. However, from an experimental point of
view, the glass transition  $T_G$ is
conventionally defined as the value of $T$ at which $\eta$ takes some large
value (this is usually taken to be $10^{13}$ poise).
In Fig. \ref{f3} we show curves of constant $\eta$ extracted from our
model. Each of these may be thought as defining a dynamical
glass transition temperature $T_G$ (that depends on pressure), according
to different dynamical criteria. The position of the line of
the fragile-to-strong transition is also indicated. 
 We see that $T_G$ is systematically lower around the critical pressure
$P=P_{\rm cr}$. This
behavior has been observed in numerical simulations of
SiO$_2$\cite{barrat,rustad}, which has a
fragile-to-strong transition qualitatively similar to that of water. 
Notice that $T_G$ is lower than the
temperature of the fragile-to-strong transition only in some range of pressure around   
$P_{\rm cr}$, and only in this range the  fragile-to-strong transition will be
experimentally observable.

One important ingredient that we have not included in the model is the
existence of an attractive part in the interaction potential. In real
water this
attraction generates the liquid-vapor first order coexistence line, and
also probably a second first order line in the supercooled region
separating two different amorphous configurations\cite{eajjcp2}. A
simple way of
analyzing the consequences on our model of an attraction between particles 
is the following. If the attraction is considered to be long ranged, of
van der
Waals type, then all the results we have obtained remain valid if we
replace
$P$ by a new effective pressure $P^*\equiv P+\gamma/v^2$ with some
constant $\gamma$, namely, the
attraction acts as an effective pressure (which depends on $v$) that
has to be added to the external pressure $P$. In the $P$-$T$ phase
diagram, this non-uniform transformation of the $P$
axis produces (if $v$ decreases rapidly when $P$ increases)
a ``folding" that indicates a first order transition\cite{eajjcp2}. This
is the way in which the liquid-vapor coexistence line appears in the van
der Waals equation for a fluid. For our
model, since $v(P)$ at $T=0$ has an abrupt discontinuity at $P=P_{\rm
cr}$, the
attraction generates also a new first order line ending in a critical point.
It seems to be\cite{stanleyrev} that in water this critical point lies at
temperatures higher than
$T_G$, so the first order line determines two different sectors of the glassy phase
of water. These two
sectors correspond to the experimentally observed high density amorphous (above the
first-order line) and low density amorphous (below the first order
line) phases\cite{stanleyrev,mishima}. In our model, these two
phases differ in the fraction of particles that are paired, and thus they
can also be named high density and low density amorphous phases.

We have relied for our discussion upon the existence of a thermodynamic
phase transition for hard spheres, that is not rigorously proven to occur.
However, for slightly different forms for $s_c^{HS}$ (which may imply the absence
of an ideal glass transition) than that given by
Eq. (\ref{schs}),
our results still remain valid, except for the fact that the sharp
fragile-to-strong transition (the kinks in the curves of Fig. \ref{f2})
becomes a crossover.

The model we have presented explains the transition between fragile and 
strong behavior of water as appearing from the competition between two
different local structures. To be able to solve the problem analytically, we had to    
make the crude approximation that
these structures correspond to
singled and paired particles.
 In real water it is likely that what plays
the role of our particles are the so-called Walrafen
pentamers\cite{walrafen}, which are
clusters of five water molecules. These cluster are
naturally expected to accommodate at one of two possible distances from
each other\cite{stanleyrev}. It is clear that in this more general case the mechanism
for a fragile-to-strong transition may remain
basically the same. In fact, from the way we solved the model, it is seen that
all what is needed is the existence
of two different competing structures, independently of the details of
them. At high
$T$ the fragile behavior is associated to the
configurational entropy of each structure individually. At low $T$ the
strong behavior  
appears due to the combinatorial entropy of choosing locally between the two
structures.

\newpage

\begin{references}


\bibitem{angell91}Angell C A 1991 {\it J. Non-Cryst. Solids} {\bf
131-133} 13
\bibitem{angell95}Angell C A 1995 {\it Science} {\bf 267} 1924
\bibitem{ito99}Ito K, Moynihan C T and Angell C A 1999 {\it Nature} {\bf
398} 492
\bibitem{sastry99}Sastry S 1999 {\it Nature} {\bf 398} 467 
\bibitem{starr99} Starr F W, Angell C A, Speedy R J  and
Stanley H E {\it LANL preprint} cond-mat\#9903451.
\bibitem{eajpre}Jagla E A 1998 {\it Phys. Rev. E} {\bf 58} 1478
\bibitem{eajjcp2}Jagla E A 1999 {\it J. Chem. Phys.} to be published
\bibitem{sadr}
Sadr-Lahijany M R, Scala A, Buldyrev S V and Stanley H E 
1998 {\it Phys. Rev. Lett} {\bf 81} 4895
\bibitem{speedy97}Speedy R J 1997 {\it J. Chem. Phys.} {\bf 107} 3222
\bibitem{stanleyrev}Mishima O and Stanley H E 1998 {\it Nature}
{\bf 396} 329 
\bibitem{campolat}Canpolat M, Starr F W, Sadr-Lahijany M R, Scala A,
Mishima O and Stanley H E 1998  {\it Chem.
Phys. Lett.} {\bf 294} 9
\bibitem{speedyhs0} Speedy R J 1993 {\it Mol. Phys.} {\bf 80} 1105
\bibitem{speedyhs1} Speedy R J 1994 {\it Mol. Phys.}
{\bf 83} 591
\bibitem{speedyhs2} Speedy R J 1998 {\it Mol. Phys.} {\bf 95}
169 
\bibitem{young}Young A P (Ed) 1998 {\it Spin Glasses and Random Fields} (Singapore:
World Scientific)
\bibitem{speedy-deb} Speedy R J and Debenedetti P G 1996 {\it Mol. Phys.}
{\bf 88} 1293
\bibitem{adam65} Adam G and Gibbs J H 1965 {\it J. Chem. Phys.} {\bf 43}
139
\bibitem{cho96}Cho C H, Singh S and Robinson G W 1996 {\it Phys. Rev.
Lett.} {\bf 76} 1651
\bibitem{barrat}Barrat J -L, Badro J and Gillet P {\it LANL preprint} cond-mat\#9612154
\bibitem{rustad}Rustad J R, Yuen D and Spera F J 1990 {\it Phys. Rev. A}
{\bf 42} 2081
\bibitem{mishima}Mishima O, Calvert L D and E. Whalley 1984 {\it Nature}
{\bf 310} 393\\  Mishima O, Takemura K and
Aoki K 1991 {\it Science} {\bf 254} 406 \\Mishima O 1994 {\it J. Chem.
Phys.} {\bf 100} 5910 
\bibitem {walrafen}Walrafen G E 1964 {\it J. Chem. Phys.} {\bf 40} 3249\\
Monosmith W B and Walrafen G E 1984 {\it J. Chem. Phys.} {\bf 81} 669
\end{references}
\end{document}